\documentclass[epj-spec,final]{svjour}
\usepackage{epsfig}
\usepackage{graphics}
\usepackage{amsbsy}
\def\ba{\begin{eqnarray}}
\def\ea{\end{eqnarray}}

\begin{document}
\titlerunning{ELRADGEN 2.0}
%\authorrunning{aaa}
\title{
ELRADGEN 2.0: Monte Carlo generator for simulation of radiative events 
in polarized elastic electron-proton scattering 
}
%\subtitle{ELRADGEN 2.0}
\author{I. Akushevich \inst{1}, 
A. Ilyichev  \inst{2}\fnmsep\thanks{\email{ily@hep.by}} \and  N. Shumeiko	  \inst{2}}
\institute{Duke University, Durham, USA \and  National
Center of Particle and High Energy Physics,
220040 Minsk, Belarus }
\abstract{
A new version of Monte Carlo generator ELRADGEN 
for simulation of real photon emission 
in elastic electron-proton scattering
is presented.  The extensions in the new version include
opportunity to deal with 
polarized particles:
longitudinally polarized electron
and arbitrary polarized proton. 
Simulation strategy, specifications of used kinematics, structure of 
the contributions to the observed cross section, 
cross-checks, and numerical results for BLAST experimental 
setup are presented and briefly discussed.
}%
\maketitle
\section{Introduction}
Exclusive real photon production in lepton-nucleon scattering
plays a rather important role in the investigation of
the nucleon structure. The measurement of
this process in different kinematical regions allows researchers to obtain the
information about the generalized parton distributions \cite{Diehl,HERMES}
and the generalized polarizabilities \cite{Gui,Dre98}.
In some cases this process appears as background to  
lepton-nucleon scattering in elastic \cite{MaxTj} and inelastic \cite{ASh}
channels.

Here we present a new version of the Monte Carlo generator
ELRADGEN. The previous one \cite{elradgen} was developed for
simulation of real hard photon emission from the lepton legs as
background to unpolarized elastic electron-proton scattering.
In the new version {\bf 2.0} we extend the generator to deal with
initial polarized particles: longitudinally polarized lepton and
arbitrary polarized proton. 
Both new and previous versions of this generator are based on the results
of ref.~\cite{AAM} which were obtained using the
Bardin-Shumeiko covariant approach
for the extraction and cancellation of the infrared divergence \cite{BSh}.

\section{Kinematics and Method of Generation}
\label{kin}

\begin{figure}[t]
\unitlength 1mm
%\hspace*{-2cm}
\vspace*{-20mm}
\begin{tabular}{cc}
\begin{picture}(80,80)
\put(-2,0){
\epsfxsize=7cm
\epsfysize=6cm
	\epsfbox{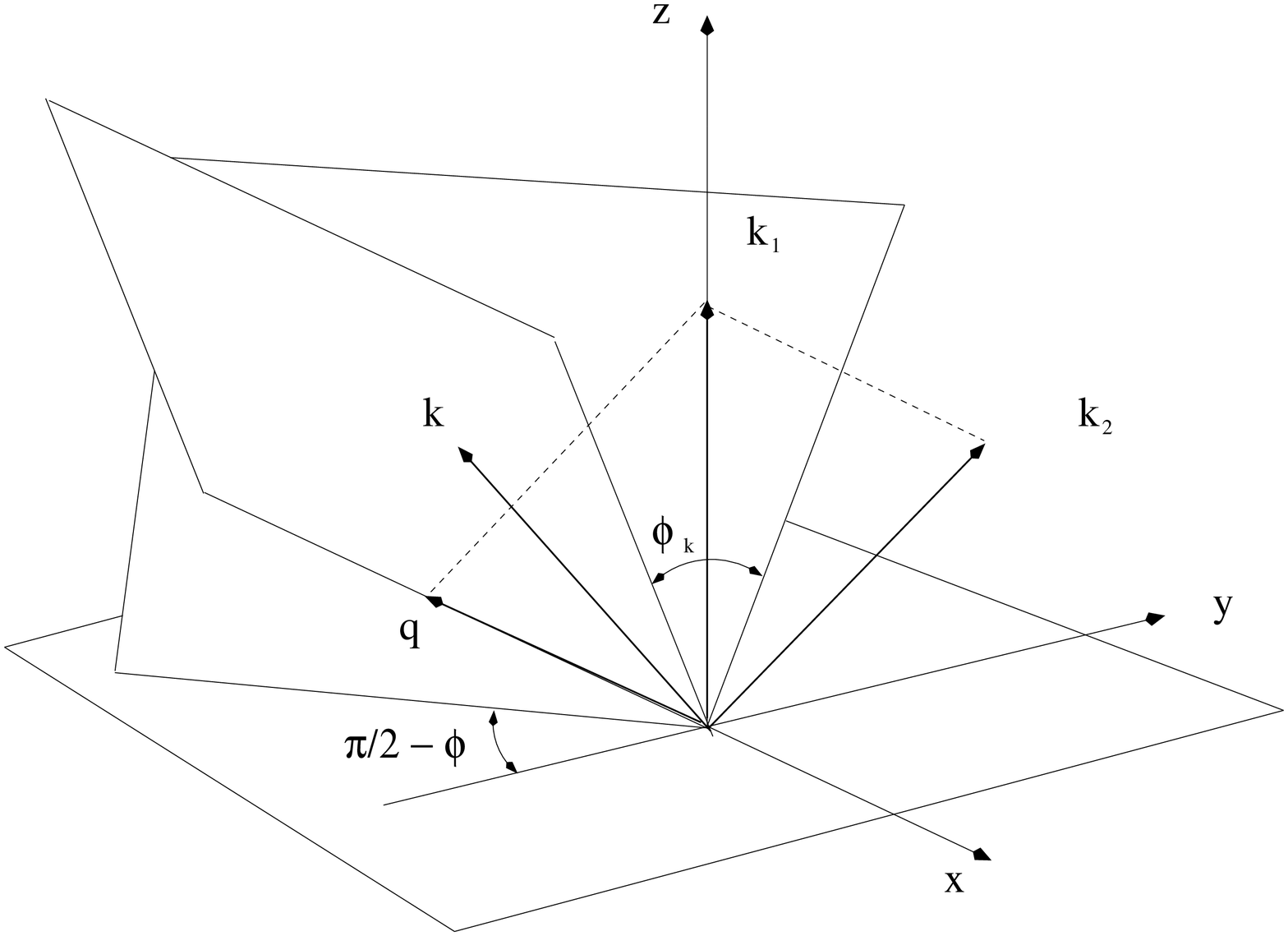}
\put(-30,-6){\mbox{(a)}}
}
\end{picture}
&
\begin{picture}(80,80)
\put(-14,0){
\epsfxsize=7cm
\epsfysize=6cm
\epsfbox{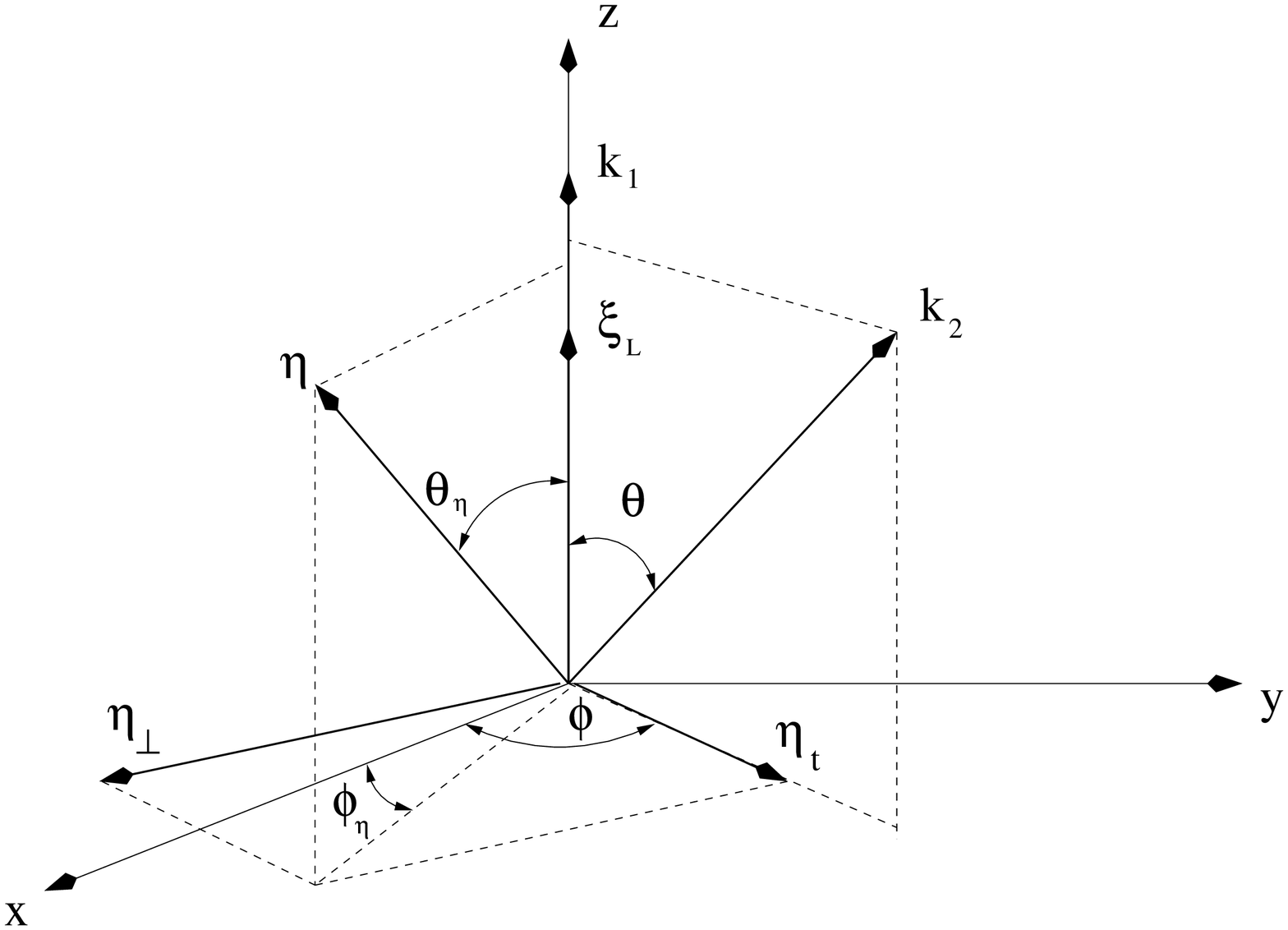}
\put(-40,-6){\mbox{(b)}}
}
\end{picture}
\end{tabular}
\vspace*{8mm}
\caption{Three-vectors decomposition in Lab system: 
a) momenta leptons and photon; b) polarized vector
for longitudinally polarized
lepton $\xi _L$ and arbitrary polarized nucleon $\eta $. 
}
\label{decv}
\end{figure}

For the simulation of exclusive radiative events in polarized electron-proton
scattering
\begin{equation}
e(k_1,\; \xi _L)+p(p_1,\; \eta )\longrightarrow e'(k_2)+p'(p_2)+\gamma (k)
\label{re}
\end{equation}
($k^2=0$, $k_1^2=k_2^2=m^2$, $p_1^2=p_2^2=M^2$)
we choose three kinematic variables: a transfer momentum squared
$t=-(k_1-k_2-k)^2$,
the inelasticity $v=(p_2+k)^2-M^2$, and the azimuthal angle $\phi_k$
between the planes ${\bf(q,k)}$ and  ${\bf
(k_1,k_2)}$ depicted in Fig.~\ref{decv}~(a).
Together with the kinematic variables characterizing Born contribution to elastic scattering
\ba
Q^2=-q^2=-(k_1-k_2)^2,\qquad S=2k_1p_1,\quad  \phi,
\label{k0}
\ea
%%%%%%%%%%%%%%%%%that are generated externally ($\phi $, naturally, flat)
they represent a full set variables for reconstruction of 
the four--vectors of all final
particles in any frame.

The four--vectors $\xi _L$ and $\eta$ in (\ref{re}) describe
the longitudinally polarized electron and
the arbitrary polarized proton, respectively.
The proton polarization is characterized by two angles $\theta _{\eta} $ 
and $\phi _{\eta }$ as presented in Fig.~\ref{decv}~(b). 
The explicit expressions for polarized vectors can be found
in \cite{ASh}.

Simulation of exclusive radiative events requires a separation of the total 
(or the radiatively corrected) 
cross section 
$\sigma _{obs}$\footnote{Here and later we define $\sigma \equiv d\sigma /dQ^2d\phi$} 
into the contribution of real hard photon emission 
$\sigma _{rad}(v_{min})$ and remaining part containing Born, soft photon, 
and additional 
virtual particle contributions $\sigma _{BSV}(v_{min})$. It can be performed 
by introducing a separation parameter, namely, the
minimum inelasticity value $v_{min}$ 
that can be associated with
missing mass square resolution of the detector
\cite{MERADGEN} when the final proton are not detected. 
The sum of these two positive parts
\begin{equation}
\sigma _{obs} =
\sigma _{rad}(v_{min})
+
\sigma _{BSV}(v_{min})
\label{sobr}
\end{equation}
does not depend on 
$v_{min}$ while  $\sigma _{BSV}(v_{min})$ and $\sigma _{rad}(v_{min})$
do. The numerical details of these dependencies are presented
in the next section. 

The explicit expressions for these two
contributions are similar to those for
$\sigma _{rad}(v_{min})$ and
$\sigma _{non-rad}(v_{min})$ from \cite{elradgen} with two exceptions: i) 
abbreviation $BVS$ is used instead of $non-rad $ for the remaining part of the cross section and ii) 
two structure functions representing the contributions of the polarized parts of the cross sections are 
additionally used. 

\begin{figure}[!t]
\unitlength 1mm
%\hspace*{-2cm}
\vspace*{5mm}	
\begin{tabular}{ccc}
\begin{picture}(40,40)
\put(-5,0){
\epsfxsize=5.4cm
\epsfysize=5.4cm
\epsfbox{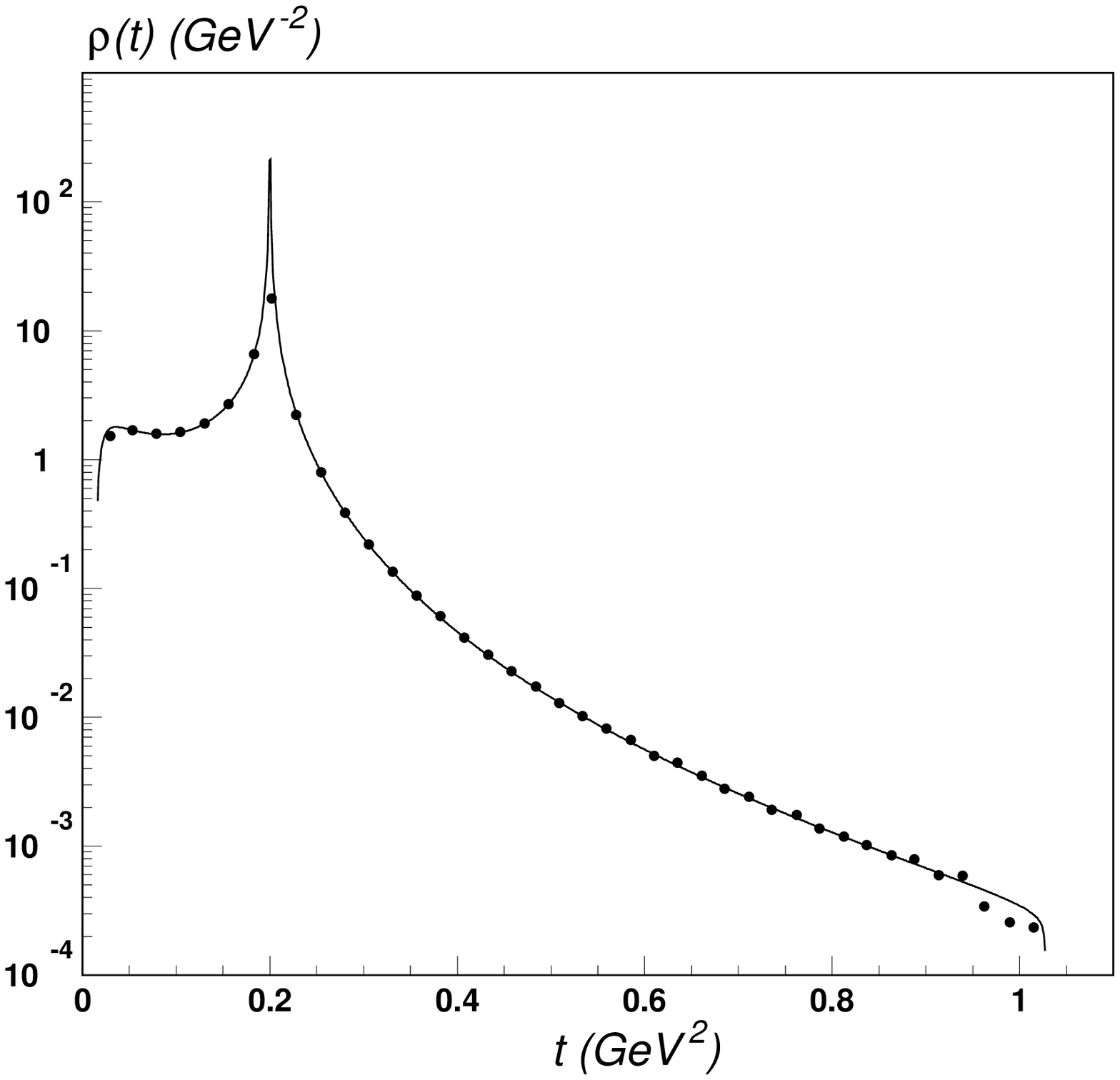}
}
\end{picture}
&
\begin{picture}(40,40)
\put(-1,0){
\epsfxsize=5.4cm
\epsfysize=5.4cm
\epsfbox{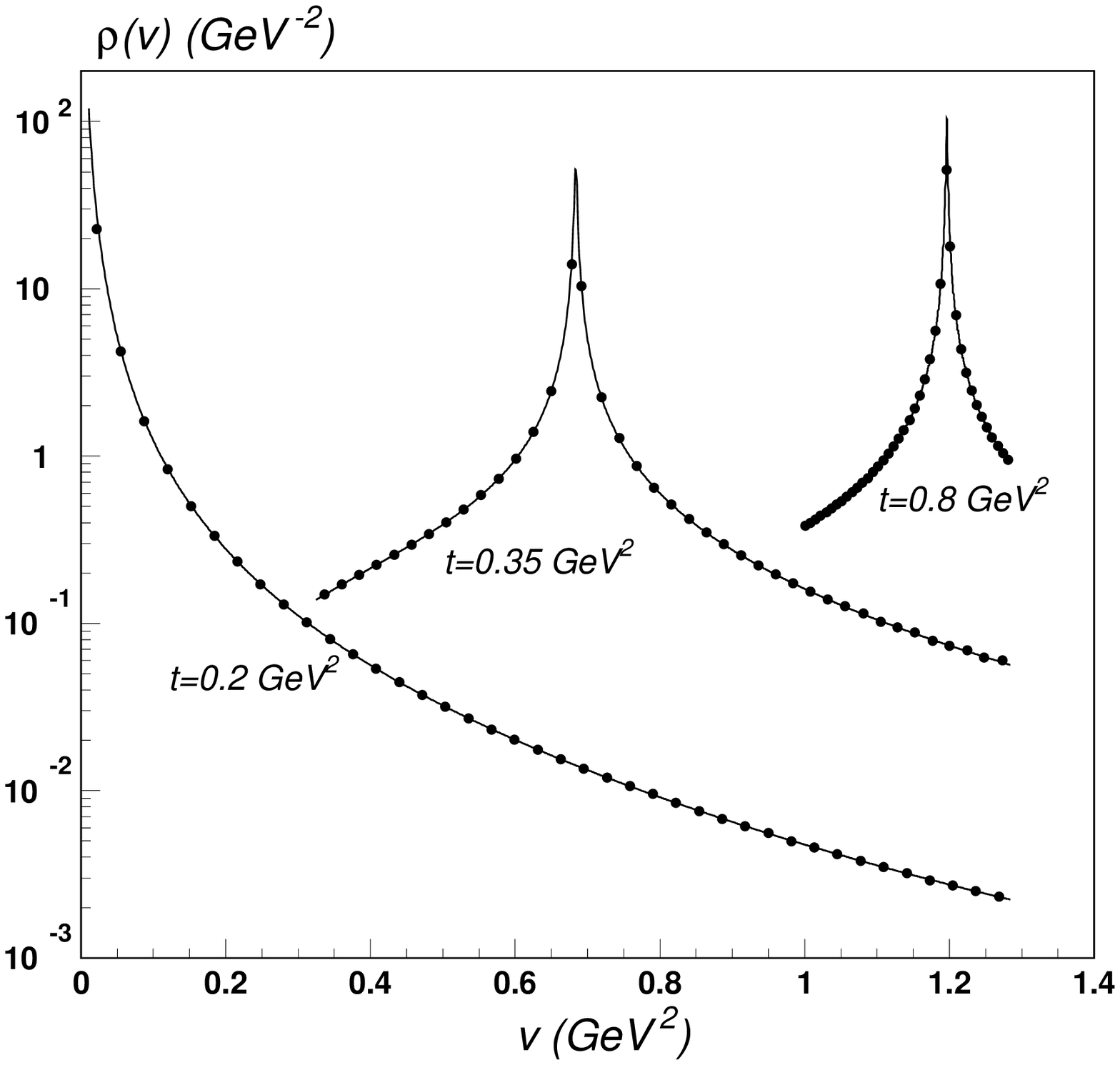}
}
\end{picture}
\begin{picture}(40,40)
\put(7,0){
\epsfxsize=5.4cm
\epsfysize=5.4cm
\epsfbox{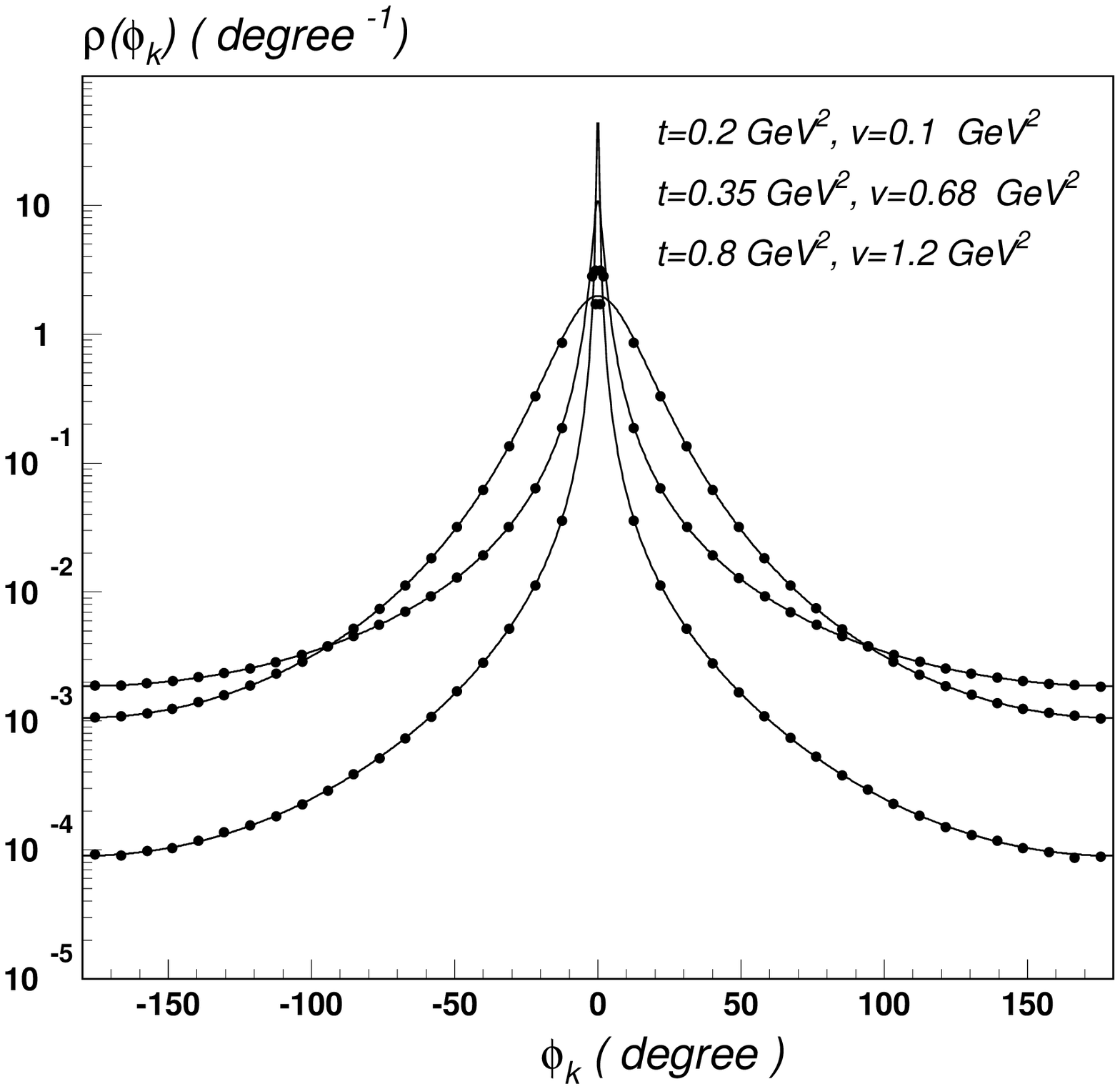}
}
\end{picture}
\end{tabular}
\vspace*{-3mm}	
\caption{Histograms (points)
and corresponding probability densities  (solid lines)
for variables describing the
 exclusive real hard photon production in polarized
electron proton scattering
at BLAST kinematic conditions \cite{BLAST}
($E_{beam}=850$~MeV, $Q^2=0.2$ GeV$^2$, 
$\theta_{\eta}=48^0$), with 
$\phi=\phi_{\eta}$, $P_LP_N=-1$
and $v_{min}=10^{-2}$~GeV$^2$. 
}
%\vspace*{3mm}	
\label{dist}
\end{figure}

 The strategy for simulation of an event can be outlined as follows.
Two contributions 
$\sigma _{rad}(v_{min})$ and $\sigma _{BSV}(v_{min})$ to 
$\sigma _{obs}$ are calculated using a predetermined value of the $v_{min}$. 
Then the channel of scattering (i.e., the process with or
without real hard photon emission) is simulated
according to the partial contributions of these two parts to
the radiatively corrected cross section.
If the channel with the real hard photon emission 
is chosen, three photonic variables
$t$, $v$ and $\phi _k$ are simulated   
according to their calculated distributions as they contribute to
$\sigma _{rad}(v_{min})$ (see Fig.\ref{dist} and ref. \cite{elradgen} for details). 
Finally using these variables together with Born ones (\ref{k0}) which can be simulated according to the Born cross section or 
be externally  predetermined, the four-vectors of all final particles in any frame are reconstructed.

The cross section of the process (\ref{re}) when
the real hard photon emitted from the lepton legs is expressed through 
the nucleon form factors which depend only on one
integration
variable, namely, on $t$.  Therefore 
to have an convenient opportunity to apply this generator to different fits or  models of 
nucleon form factors, the
integration over $t$ has to be used numerical only, 
while the analytical integration over the other 
photonic variables $v$ and $\phi _k$ is possible and 
allows to speed up the process of event
generation. The analytical integration over these two variables was used 
in previous version of this generator for unpolarized scattering  \cite{elradgen}. 
 However when we deal with  the arbitrary
polarized proton the analytical integration over $v$
is a rather difficult due to non-trivial dependence 
of transverse component of the proton polarized vector
on this variable.
Therefore, in present version of the generator the analytical integration over $v$ is used  
for unpolarized particle scattering only.  

\section{Numerical test}
\label{check}
One cross-check which has to be done first is to investigate how well the simulated 
distributions of photonic variables $t$, $v$ and $\phi _k$ reproduce those theoretically calculated.   
Fig. \ref{dist} provides such an illustration for 
the polarized electron proton scattering at BLAST kinematic 
conditions \cite{BLAST}: the longitudinally polarized
electrons with energy 850 MeV scatter off the proton that 
polarized under angle
$\theta _{\eta }=48^0$ to the lepton beam. Plots 
for the
photonic variable distributions contributing to $\sigma_{rad} (v_{min})$
demonstrate excellent quality of the simulation procedure. 
Here $P_L$ ($P_N$)
is the degree of electron (proton) polarization,
the solid lines corresponds to probability densities  
calculated numerically and the points show 
the event distributions simulated by ELRADGEN.
Nucleon form factors are taken from \cite{ff}.
In the left plot the theoretical $t$-distribution is presented 
when the integration over
two  other photonic variables has been performed. The sharp peak corresponds the situation
when $t=Q^2$. On the second plot one can see $v$-distributions
at the fixed $t$ and integrated $\phi _k$ variables. Here the peak at $t=0.2$ GeV$^2$ near 
small $v$ corresponds the infrared divergence (that cut off by $v_{min}$). 
The peaks on other two distributions correspond the situation when 
the real photon emitted along the momentum of scattered lepton. 
The right plot presents $\phi _k$-distribution for fully differential
contribution of real hard photon emission. As one can see,
the momenta of the most of emitted photons are concentrated near scattering plane i.e. when
$\phi_k=0$. 

Two studies were illustrated in Tab.~\ref{tab:2} :
i) 
the investigation of  the $v_{min}$-dependence of
$\sigma _{rad}(v_{min})$, 
$\sigma _{BSV}(v_{min})$ and their sum;
and 
ii) 
the comparison
of the ratio of the radiatively corrected cross section $\sigma _{obs}$ to the 
Born contribution  $\sigma _{0}$ obtained by our generator and 
Fortran code MASCARAD \cite{AAM} with the same input parameters as a simplest comparison
of these two codes. 
Specifically,  Tab.~\ref{tab:2} demonstrates that 
the observable cross section does not almost change 
with decreasing $v_{min}$ from 1
to $10^{-4}$ GeV$^2$ 
while its components $\sigma _{rad}(v_{min})$ and
$\sigma _{BSV}(v_{min})$ change essentially:
$\sigma_{rad} (v_{min} )$ increases and $\sigma_{BSV} (v_{min})$ decreases. 
The comparison with MASCARAD results in a good agreement as well.

\begin{table}
\caption{The $v_{min}$-dependence of the radiative, BSV
and observable contributions to  electron-proton
scattering with polarized target 
for different spin orientation
in the Born units and results of comparison with MASCARAD \cite{AAM} 
at BLAST kinematic conditions \cite{BLAST}
($E_{beam}=850$~MeV, $Q^2=0.2$ GeV$^2$, 
$\theta_{\eta}=48^0$), with 
$\phi=\phi_{\eta}$.
}
\label{tab:2}
\centering{
\begin{tabular}{|c|c|c|c|c|c|c|c|c|}
\hline 
$v_{min}$ 
&
\multicolumn{2}{c|}{$\sigma _{rad}/\sigma_0$}
&
\multicolumn{2}{c|}{$\sigma _{BSV}/\sigma_0$}  
&
\multicolumn{4}{c|}{$\sigma _{obs}/\sigma_0$}   
\\\cline{2-9}
GeV$^2$
&
\multicolumn{2}{c|}{ELRADGEN}
&
\multicolumn{2}{c|}{ELRADGEN}  
&
\multicolumn{2}{c|}{ELRADGEN}   
&
\multicolumn{2}{c|}{MASCARAD}   
\\\hline
$P_LP_N$
&1
&-1
&1
&-1
&1
&-1
&1
&-1
\\ 
\hline
1	              &0.01562&0.02342& 1.018& 1.030&1.033&1.053&&\\ 
\cline{1-7}$10^{-1}$&0.1299 &0.1484 &0.9029&0.9040&1.033&1.052&&\\ 
\cline{1-7}$10^{-2}$&0.2580 &0.2776 &0.7726&0.7727&1.031&1.050&1.033&1.053\\ 
\cline{1-7}$10^{-3}$&0.3873 &0.4070 &0.6388&0.6388&1.026&1.046&&\\
\cline{1-7}$10^{-4}$&0.5192 &0.5389 &0.5046&0.5046&1.024&1.043&&\\ \hline
\end{tabular}}
%\vspace*{5mm}
\end{table}
\section{Conclusion and Outlook}  
\label{cconc}
In the present report the new version of the Monte Carlo generator ELRADGEN for
simulation of real photon events within
elastic electron-proton scattering generalized
for longitudinally polarized lepton and
arbitrary polarized target is presented.
   
Numerical test of new version of this code shows a good agreement with the Fortran
code MASCARAD \cite{AAM} and reveals
lack of dependence on minimum inelasticity value $v_{min}$  
with accuracy up to 1\%. Besides we found that 
the distributions of generated radiative events are
in coincidence with corresponding 
probability density.

The present approach is rather general and can be extended
in many other different ways including i) 
the development of this generator for transferred
polarization from lepton beam
to recoil proton \cite{AhR}
for 
measurement of electromagnetic form-factors of the proton
in polarized scattering \cite{FFpol1,FFpol2};
ii) 
its further generalization for the investigation 
of electroweak effects such as axial
form factors of the nucleon \cite{Gor} and
parity violation elastic scattering \cite{Beck};
and iii)
its generalization for practical involvement in the experiments
with the measurement of 
generalized parton distribution \cite{Diehl,HERMES}
as well as 
generalized polarizabilities
\cite{Gui,Dre98}.
%\end{itemize}

\begin{acknowledgement}
We would like to acknowledge useful discussion with E. A. Kuraev. 
The one of us (A. I.) would like to thank the staff of MIT Bates Center
as well as O. F. Filoti for
their generous hospitality during his visit.
\end{acknowledgement}

\end{document}